\documentclass[a4paper,11pt]{article}
\usepackage{jheppub} 
\usepackage{lineno}
\usepackage{float}

\usepackage{amsmath}
\usepackage{amssymb}
\usepackage{amsfonts}
\usepackage{dsfont}
\usepackage{xcolor}
\usepackage{comment}
\usepackage{csquotes} 
\MakeOuterQuote{"}  

\newcommand{\ie}{\textit{i.e.}}
\newcommand{\eg}{\textit{e.g.}}
\newcommand{\h}{\hspace{0.5mm}}

\title{\boldmath Confinement and chiral symmetry breaking in holography: a smooth switch-off}

\author[a]{Martí Berenguer,}
\author[b]{Johanna Erdmenger,}
\author[c]{Nick Evans,}
\author[c]{Wanxiang Fan,}
\author[b]{Florian Vasel}

\affiliation[a]{Departamento de Física de Partículas, Universidade de Santiago de Compostela and Instituto Galego de Física de Altas Enerxías (IGFAE). E-15782 Santiago de Compostela, Spain.}
\affiliation[b]{Institute for Theoretical Physics and Astrophysics, Julius Maximilians University W\"urzburg, Am Hubland, DE-97074 W\"urzburg, Germany.}
\affiliation[c]{School of Physics \& Astronomy and STAG Research Centre, University of Southampton,
Highfield, Southampton SO17 1BJ, UK.}

\emailAdd{marti.berenguer.mimo@usc.es}
\emailAdd{evans@soton.ac.uk}
\emailAdd{w.fan@soton.ac.uk}
\emailAdd{florian.vasel@stud-mail.uni-wuerzburg.de}

\abstract{
We revisit the holographic description of the thermal first order phase transition of $\mathcal{N}=4$ SYM compactified on a spatial circle. At the transition, the dominant bulk saddle exchanges between a geometry with a compact spatial circle and one with a compact Euclidean time circle. We construct a one-parameter family of Euclidean geometries that describes the unstable branch of the transition, completing the swallow-tail structure of the free energy. Although these configurations are thermodynamically unstable, they provide a continuous interpolation between the confining soliton and the deconfined black hole phases. Using probe fundamental strings, we show that the theory remains confining along the unstable branch, with a string tension that decreases smoothly and vanishes only in the black hole limit. Introducing fundamental matter via probe D5-branes, we find that chiral symmetry breaking follows the same pattern: the condensate decreases continuously and switches off precisely where confinement disappears. We discuss the implications for the confinement and chiral symmetry breaking mechanisms at large $N_c$.}

\usepackage[T1]{fontenc}
\usepackage{lmodern}
\begin{document}
\maketitle
\flushbottom

\section{Introduction} \label{sec:intro}

The precise mechanisms responsible for confinement and chiral symmetry breaking in QCD remain poorly understood. Most likely confinement is due to the condensation of magnetically charged, non-perturbative configurations in the QCD vacuum and a dual Meissner effect \cite{tHooft:1977nqb,
Mandelstam:1974pi}. Explicit realizations of this mechanism have been identified in certain controlled setups. Notably, Yang-Mills theory with adjoint matter compactified on a circle provides a weakly-coupled example in which confinement is driven by magnetically charged instantons in the efecctive lower-dimensional IR theory \cite{Unsal:2007jx,Unsal:2007vu,Poppitz:2021cxe}. Similarly, the same mechanism also appears in the ${\cal N}=2$ super-Yang Mills (SYM) theory broken to ${\cal N}=1$, as described in the Seiberg-Witten theory \cite{Seiberg:1994rs}.

Bag-type models suggest  that bound state masses will develop using an argument based on the uncertainty principle \cite{Chodos:1974je,Bars:1982zq}.  The smaller the box that particles are confined to, the larger their average momenta and hence the energy.  Whilst it is not a proof that the mass gap results from a chiral symmetry breaking quark condensate, here we will work in a holographic model which does make this link. Confinement is not a necessary condition for chiral symmetry breaking though. For example, an external magnetic field in a fermionic system can trigger chiral symmetry breaking even at weak coupling \cite{Miransky:2015ava}. The trigger for chiral symmetry breaking has been argued to occur when the anomalous dimension of the quark mass operator reaches $\gamma=1$, and the dimension of the mass and condensate become equal \cite{Miransky:1984ef,Cohen:1988sq}. In principle, this criterion could be realized in QCD before confinement sets in. The scales would necessarily be close because the quarks would condense at strong values of the QCD coupling that also runs fast. In \cite{Evans:2020ztq,Alfano:2024aek,Alfano:2025dch} theories where this gap might be larger have been proposed. 

Holography is now a well-established technique for studying strongly coupled gauge theories and is believed to yield exact answers in the large-$N_c$ limit for ${\cal N}=4$ SYM and its deformations. Early work already demonstrated that placing the theory on a compact manifold can lead to confinement \cite{Witten:1998zw}, and that increasing the temperature induces a first-order deconfinement transition to a black hole geometry. Despite this success, the microscopic origin of confinement in holography remains obscure. One reason is that the gravitational description captures only gauge-invariant degrees of freedom, so monopoles will not be explicitly visible. Moreover, the discontinuous first order nature of the transition makes it difficult to track how confinement is switched on or off.   

In this paper we wish to improve our understanding of confinement in holographic gauge theories, and in particular, its relation to chiral symmetry breaking. Our goal is to find a controlled setting in which both phenomena can be turned off smoothly within a fully top-down construction, without introducing phenomenological input.

Ideally, one would like to study a theory in which confinement and chiral symmetry breaking disappear through a second-order transition, allowing for a smooth interpolation between phases. Here though we take a more circuitous route by exploiting a generic feature of first-order transitions: the presence of unstable saddle points that complete the swallow-tail structure of the free energy. For a range of parameters near the transition, the free energy typically admits three extrema: two local minima corresponding to the competing stable phases, and an unstable local maximum interpolating between them. Although this intermediate saddle is thermodynamically unstable and contains tachyonic modes, these instabilities will be in the gauge sector of the theory and the geometry will provide a continuous path connecting the two phases. Our central idea is to use this unstable branch as a controlled interpolation between confining and deconfining physics.

We implement this idea in the well-known example of ${\cal N}=4$ SYM compactified on a spatial circle of radius $R_z$ \cite{Aharony:2006da}.  The introduction of this scale breaks conformal invariance and leads, at zero temperature, to the AdS soliton geometry \cite{Horowitz:1998ha}, which exhibits confinement: Wilson loops for widely separated quarks are described by strings that lie along the capped-off IR region, producing a linear potential \cite{Maldacena:1998im}. Finite temperature can be introduced via a compact time direction, leading to a competing AdS black hole geometry, and at sufficiently high temperature it dominates the free energy. In this phase, widely separated quarks lead to strings that fall into the horizon and the quarks become screened from each other \cite{Brandhuber:1998bs}. The transition between these two geometries is first order and has been extensively studied \cite{Aharony:2006da}.

To study the physics along the unstable branch of this transition, we construct a one-parameter family of Euclidean geometries that interpolate continuously between the soliton and the black hole. These geometries correspond to different bulk fillings of the same boundary two-torus spanned by Euclidean time $\tau$ and the compact spatial direction $z$. The interpolating configurations arise as rotations of the soliton and black hole metrics in the $\tau$–$z$ plane, and realize the unstable extrema that complete the swallow-tail structure of the free energy.

Fundamental matter can be introduced into the theory via probe D5-branes, which place the quarks on a 2+1 dimensional defect. In the soliton background, quarks with vanishing ultraviolet mass dynamically acquire an infrared mass, signaling chiral symmetry breaking, while in the black hole background this does not occur. 

We then study confinement and chiral symmetry breaking along the unstable branch by analyzing probe strings and D5 brane embeddings in this family of geometries. A key feature is that a true event horizon exists only in the strict black hole limit. Away from this limit, probes that would normally fall through the horizon instead skirt a near-horizon region, with their action vanishing smoothly as the black hole endpoint is approached. Consequently, both the string tension and the chiral condensate decrease continuously along the unstable branch and vanish only at the deconfined black hole phase.

The paper is organized as follows: In Section \ref{sec:swallow} we present the three geometries that we consider, and we compute the free energy of the interpolating geometry, completing the swallow-tail structure. In Section \ref{sec:confinement} we study the presence or absence of confinement in our geometry. After that, we introduce probe D5 branes to study chiral symmetry breaking in Section \ref{sec:cSB}. In the conclusions we will seek to draw lessons on  how the holographic geometry encodes the monopole density (through an expectation value of the $T^{zz}$ element of the stress energy tensor). There are many interesting aspects to our analysis and we believe it is a useful step in the study of the origin of confinement and chiral symmetry breaking.

\section{The thermal transition and its swallow tail} \label{sec:swallow}

We now describe the holographic geometries relevant to the thermal transition of ${\cal N}=4$ SYM compactified on a spatial circle, and show how the unstable branch completing the swallow tail arises.

Throughout this section we work in Euclidean signature. The boundary theory lives on a two-torus spanned by Euclidean time $\tau$ and the compact spatial direction $z$, with periods $\beta$ and $2\pi R_z$, respectively. Different bulk saddles correspond to different ways in which one of these cycles becomes contractible in the interior.

\subsection{Geometries}\label{subsec:geometries}

\paragraph{The soliton geometry.}

\begin{figure}[h!]
    \centering
    \includegraphics[width=0.45\textwidth]{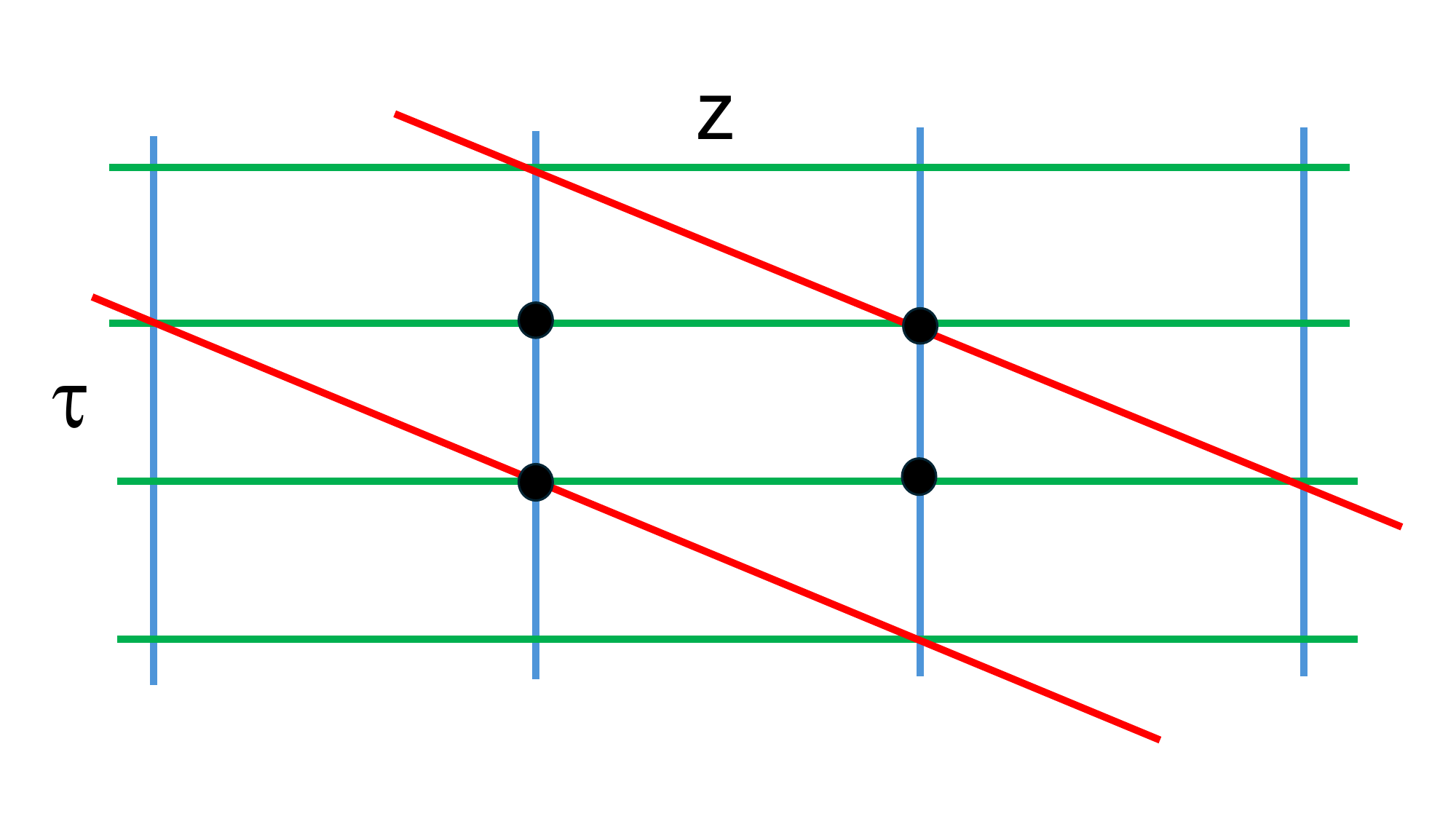}   \includegraphics[width=0.45\textwidth]{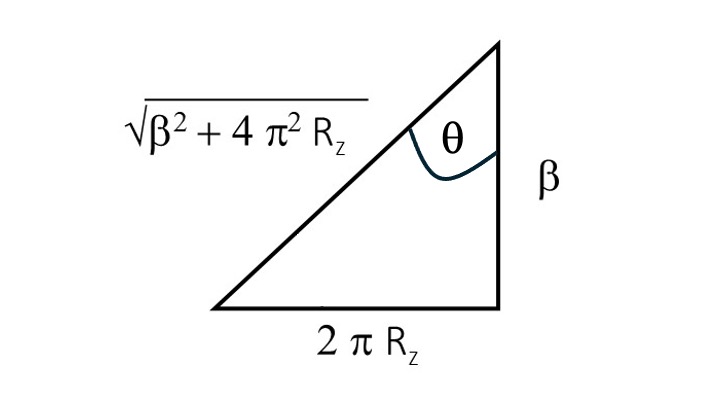} 
    \caption{We show the $\tau-z$ torus on the left as a repeating periodic rectangle.   The blue and green lines show identification in the soliton and BH geometries. One of the identifications is always enforced by the metric. The black circles are therefore all the same point. The red lines show the explicitly enforced periodicity of the interpolating metric.
    } \label{fig:triangle}
\end{figure}

At zero temperature the theory is described by the AdS soliton geometry
\begin{equation}
    ds^2=\frac{r^2}{L^2}\left(d\tau^2+dx^2+dy^2+f(r)dz^2\right)+\frac{L^2}{r^2f(r)}dr^2+L^2d\Omega_5^2~,
    \label{eq:solitonmetric}
\end{equation}
where
\begin{equation}
    f(r)=1-\frac{r_0^4}{r^4}~.
    \label{eq:BHfactor}
\end{equation}
The spacetime caps off smoothly at $r=r_0$, reflecting the compactness of the $z$ direction. Introducing coordinates $\rho^2 = \frac{L^2}{r_0} r$ and $z' = \frac{2r_0}{L^2}z$, the $(r,z)$ subspace becomes
\begin{equation}
    ds^2 = d \rho^2 + \rho^2 d z'^{2}~,
\end{equation}
so regularity fixes the period of $z$ to
\begin{equation}
    2 \pi R_z = \frac{\pi L^2}{r_0}~.
\end{equation}
The soliton geometry explicitly displays the $z$ periodicity in the metric,
\begin{equation}
    (\tau,z)\sim(\tau,z+2\pi R_z),
\end{equation}
while periodicity in $\tau$ with period $\beta$ is imposed on fluctuations to describe the finite-temperature theory. The space of the finite $T$ theory is a torus but can be considered also as a repeated rectangular region as shown in Figure~\ref{fig:triangle}. The blue lines are equivalent by the explicit $z$ periodicity of the metric whilst the green lines are equivalent by the $\tau$ periodicity imposed on solutions. The metric (and hence the vacuum) are $\tau$ invariant so naturally already have this periodicity. 

\paragraph{The black hole geometry.}

An alternative bulk filling of the same boundary torus is the Euclidean AdS black hole,
\begin{equation}
    ds^2=\frac{r^2}{L^2}\left(f(r)d\tau^2+dx^2+dy^2+dz^2\right)+\frac{L^2}{r^2f(r)}dr^2+L^2d\Omega_5^2~.
    \label{eq:BHmetric}
\end{equation}
with again
\begin{equation}
    f(r)=1-\frac{r_0^4}{r^4}~.
\end{equation}
In this case the $\tau$ circle shrinks smoothly at $r=r_0$, where an event horizon develops, and regularity fixes the Hawking temperature to be
\begin{equation}
    T_{BH}=\frac{r_0}{\pi L^2}~.
\end{equation}
The $\tau$ periodicity is explicit in the metric,
\begin{equation}
    (\tau,z)\sim(\tau+\beta,z),
\end{equation}
while the compactness of $z$ must again be imposed by hand, then describing ${\cal N}=4$ SYM at finite temperature on a compact $z$. At high temperature this geometry dominates the free energy and describes the deconfined phase.

\paragraph{Interpolating (unstable) geometries.}

The soliton and black hole correspond to two distinct ways of filling the same boundary $\tau$–$z$ torus. On top of these two familiar geometries we now wish to add a more eccentric one. As illustrated in Figure~\ref{fig:triangle}, the four corners of the fundamental rectangle are equivalent points either by one or both periodicities. One can therefore include a geometry where the single periodicity across the diagonal in $z$-$\tau$ space is manifest in the geometry. We identify points on the red lines. The other periodicity (across the opposite diagonal) must be imposed on the solutions in the space by hand.

Concretely, these geometries are obtained by rotating the soliton or black hole metrics in the $\tau$-$z$ plane by an angle $\theta$,
\begin{equation}
    \tau=\tau' \cos \theta - z' \sin \theta~,\qquad z= z' \cos \theta + \tau' \sin \theta
    \label{eq:rotation}
\end{equation}
with $\theta=0$ corresponding to the black hole and $\theta=\pi/2$ to the soliton. Denoting by $x_1$ the diagonal in the $z$-$\tau$ space, we note that regularity fixes the periodicity of the contractible cycle to
\begin{equation}
    \Delta x_1= \frac{\pi L^2}{r_0}=\sqrt{\beta^2 + 4 \pi^2 R_z^2}~,
    \label{eq:Deltax1}
\end{equation}
while the periodicity along the opposite diagonal is given by
\begin{equation}
    \Delta x_2=\beta\sin\theta = \frac{2\pi R_z\beta}{\sqrt{\beta^2 + 4 \pi^2 R_z^2}}.
    \label{eq:Deltax2}
\end{equation}
The $\theta=0$ (black hole) limit occurs in the zero-temperature limit, while the geometry coincides with the soliton at $\theta=\pi/2$ or infinite temperature.

Performing the rotation \eqref{eq:rotation} explicitly, the interpolating metric is
\begin{equation}
\begin{aligned}
    ds^2=&\frac{r^2}{L^2}\left(\sin^2\theta+f(r)\cos^2\theta\right)d\tau'^2+\frac{r^2}{L^2}(dx^2+dy^2)+(f(r)-1)\frac{r^2}{L^2}\sin 2\theta ~d\tau'dz'\\
    +&\frac{r^2}{L^2}\left(\cos^2\theta+f(r)\sin^2\theta \right)dz'^2+\frac{L^2}{r^2f(r)}dr^2+L^2d\Omega_5^2~,
\end{aligned} \label{eq:intermetric}
\end{equation}
or writing the blackening factor explicitly,
\begin{equation}
\begin{aligned}
    ds^2=&\frac{r^2}{L^2}\left(1-\frac{r_0^4}{r^4}\cos^2\theta\right)d\tau'^2+\frac{r_0^4}{L^2r^2}\sin(2\theta) d\tau'dz'+\frac{r^2}{L^2}(dx^2+dy^2)\\
    +&\frac{r^2}{L^2}\left(1-\frac{r_0^4}{r^4}\sin^2\theta\right)dz'^2+\frac{L^2}{r^2f(r)}dr^2+L^2d\Omega_5^2~.
\end{aligned}
\end{equation}
These geometries satisfy the identification
\begin{equation}
    (\tau,z)\sim(\tau+\beta,z+2\pi R_z)~.
\end{equation}
We must be careful to compare three solutions with the same torus at the boundary ie with the same $\Delta \tau, \Delta z$.

\subsection{Thermodynamics}

To compute the phase diagram of the theory we need to compute the free energy of \eqref{eq:intermetric} as a function of $\beta, R_z$ \cite{Hawking:1982dh}.  To find the free energy for all our geometries it is sufficient to consider the generic Euclidean metric
\begin{equation}
    ds^2=\frac{r^2}{L^2}\left[\left(1-\frac{r_0^4}{r^4}\right)dx_1^2+dx_2^2+\sum_{i=3}^4dx_i^2\right]+\frac{L^2}{r^2f(r)}dr^2~.
    \label{eq:generalEuclidean}
\end{equation}
The coordinates $x_1$ and $x_2$ are periodic with periods $\Delta x_1$ and $\Delta x_2$, respectively. The periodicity of $x_1$ is fixed by requiring regularity of the metric in the IR, \ie ~\eqref{eq:Deltax1}, while the periodicity of $x_2$ is imposed by hand and is given by \eqref{eq:Deltax2}. These periodicities are chosen in such a way that we always sit on the base rectangle of Figure~\ref{fig:triangle} . 

Different choices of identifications correspond to the various geometries of interest. In particular, the Euclidean black hole is obtained by identifying $x_1$ with the Euclidean time direction, which gives the contractible circle in the bulk, while $x_2$ is the compact spatial direction with radius $2\pi R_z$. Conversely, the soliton geometry is obtained by identifying $x_1$ with the spatial direction that shrinks in the bulk, while $x_2$ is the compactified Euclidean time with period $\beta$.

Although the periods $\Delta x_1$, $\Delta x_2$ satisfy $\Delta x_1\Delta x_2=2\pi R_z\beta$ in all cases, we will keep $\Delta x_1$ and $\Delta x_2$ explicit in the following for clarity.

\begin{figure}
    \centering
    \includegraphics[width=0.5\textwidth]{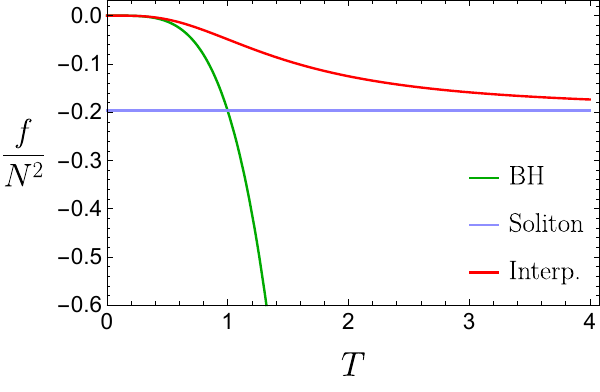}
    \caption{The free energy of the metrics \eqref{eq:solitonmetric}, \eqref{eq:BHmetric} \eqref{eq:intermetric} as a function of the temperature for $R_z=1/2\pi$, leading to the well-known first-order phase transition at $T=\frac{1}{2\pi R_z}$. The colors used are consistent with the identifications in Fig. \ref{fig:triangle}.}
    \label{fig:freeenergy}
\end{figure}

For the evaluation of the free energy, it is convenient to change variables to $\tilde{z}\equiv L^2/r$, and the metric becomes
\begin{equation}
    ds^2=\frac{L^2}{\tilde{z}^2}\left[\left(1-\frac{\tilde{z}^4}{\tilde{z}_0^4} \right)dx_1^2+dx_2^2+\sum_{i=3}^{4}dx_i^2+\frac{d\tilde{z}^2}{1-\frac{\tilde{z}^4}{\tilde{z}_0^4} }\right]
\end{equation}
with $\tilde{z}_0\equiv L^2/r_0$. The free energy $F$ is given by $\beta F=S_E$, where $S_E$ is the Euclidean action,
\begin{equation}
    S_E=-\frac{1}{16\pi G_5}\int d^5x\sqrt{g}(R-2\Lambda)-\frac{1}{8\pi G_5}\int d^4x\sqrt{h}K~.
\end{equation}
In our case, $R=-20/L^2$, $\Lambda=-6/L^2$. 
After introducing a UV cutoff $\epsilon \ll 1$, the Einstein-Hilbert term is
\begin{equation}
    S_{EH}=-\frac{1}{16\pi G_5}\int d^5x\sqrt{g}(R-2\Lambda)=-\frac{L^3V_2}{8\pi G_5}\Delta x_1\Delta x_2\left(-\frac{1}{\epsilon^4}+\frac{1}{\tilde{z}_0^4}\right)
\end{equation}
For the Gibbons-Hawking term we need the induced metric at the boundary:
\begin{equation}
    ds^2=\frac{L^2}{\epsilon^2}\left[\left(1-\frac{\epsilon^4}{\tilde{z}_0^4}\right)dx_1^3+dx_2^2+\sum_{i=3}^{4}dx_i^2\right]
\end{equation}
which has $\sqrt{h}=\frac{L^4}{\epsilon^4}\sqrt{1-\frac{\epsilon^4}{\tilde{z}_0^4}}\simeq \frac{L^4}{\epsilon^4}-\frac{L^4}{2\tilde{z}_0^4}$. We can use $\sqrt{h}K=n^{\mu}\partial_{\mu}\sqrt{h}$ with the only non-vanishing component of the normal vector being $n^{\tilde{z}}=-\frac{\tilde{z}}{L}\sqrt{1-\frac{\tilde{z}^4}{\tilde{z}_0^4}}$. We obtain
\begin{equation}
    \sqrt{h}K=\frac{4L^3}{\epsilon^4}-\frac{2L^3}{\tilde{z}_0^4}
\end{equation}
and therefore the Gibbons-Hawking term is
\begin{equation}
    S_{GH}=-\frac{1}{8\pi G_5}\int d^4x\left(\frac{4L^3}{\epsilon^4}-\frac{2L^3}{\tilde{z}_0^4}\right)=-\frac{L^3V_2}{8\pi G_5}\Delta x_1\Delta x_2 \left(\frac{4}{\epsilon^4}-\frac{2}{\tilde{z}_0^4}\right)
\end{equation}
Adding both terms we obtain the regularized on-shell action,
\begin{equation}
    S_{reg}=S_{EH}+S_{GH}=-\frac{L^3V_2}{8\pi G_5}\Delta x_1\Delta x_2\left(\frac{3}{\epsilon^4}-\frac{1}{\tilde{z}_0^4}\right)
\end{equation}
This requires the usual counterterm corresponding to the infinite volume of AdS,
\begin{equation}
    S_{ct}=\frac{3}{8\pi G_5 L}\int d^4x\sqrt{h}=\frac{L^3 V_2}{8\pi G_5}\Delta x_1\Delta x_2\left(\frac{3}{\epsilon^4}-\frac{3}{2\tilde{z}_0^4}\right)
\end{equation}
The renormalized action is then
\begin{equation}\label{eq:ren_action}
S_{ren}=\lim_{\epsilon\rightarrow 0}\left(S_{reg}+S_{ct}\right)=-\frac{L^3V_2}{16\pi G_5}\frac{\Delta x_1\Delta x_2}{\tilde{z}_0^4}
\end{equation}
Now we can reintroduce $G_5=\frac{\pi L^3}{2N^2}$ and $\pi \tilde{z}_0 = \sqrt{\beta^2+4\pi^2 R_z^2}$. The free energy is just $F=T S_E$:
\begin{equation}
    f\equiv \frac{F}{V_2}=-\frac{N^2}{4}\frac{\pi^3R_zT^4}{\left(1+4\pi^2R_z^2T^2\right)^2}
\end{equation}
For small (large) $R_z T$, we recover the well-known BH (soliton) free energies \cite{Burgess:1999vb}
\begin{equation}
\begin{aligned}
    f_{BH}&=-\frac{N^2}{4}\pi^3R_zT^4\\
    f_{sol}&=-\frac{N^2}{64\pi R_z^3}~,
\end{aligned}
\end{equation}
which are understood as relative to the pure $AdS$ vacuum chosen as reference background. The three solutions for the free energy are displayed in Fig. \ref{fig:freeenergy} for fixed $R_z=1/2\pi$. As expected, the system exhibits a first-order phase transition between the low-temperature soliton phase and the high-temperature black hole phase. The interpolating solution always has a higher free energy than either of these phases, corresponding to the local maximum of the effective potential between the two minima that interchange. The interpolating geometry completes the expected swallow tail structure associated with any first order transition. The reader should note that the natural ordering of geometries with respect to $T$ on the unstable branch of the swallow tail is the opposite of that for the physical solutions. For the physical solutions, the soliton geometry dominates at $T=0$ whilst the black hole dominates at large $T$. On the unstable branch, the black hole-like solution is present at $T=0$ and it smoothly maps to the soliton at large $T$. Whilst un-intuitive, this is the naturally expected behavior around the swallow tail.

While the completion of the swallow tail is interesting, the main reason we are interested in the extra solutions is that they provide a smooth (albeit metastable) interpolation between the soliton and black hole solutions.

\section{Confinement/deconfinement transition}\label{sec:confinement}

The presence or absence of confinement can be diagnosed by including open strings in the geometry whose endpoints are anchored at the boundary and represent infinitely heavy external quarks \cite{Maldacena:1998im,Brandhuber:1998bs}. We consider a quark–antiquark pair separated along one of the non-compact spatial directions, which we take to be $x$ (but not the compact $z$ direction). Note that in the interpolating geometry (\ref{eq:intermetric}) the off diagonal terms vanish on the boundary so the usual identification for the coordinates of the Wilson loop remains sensible. Thus we will be generically interested in a string with a profile $x(r)$. For interacting quarks the string dips into the bulk of AdS reaching a minimum radial value of $r_{min}$.

The dynamics of the string is governed by the Nambu–Goto action evaluated in the interpolating geometry \eqref{eq:intermetric},
\begin{equation}
\begin{aligned}
    S_{NG}&=\frac{1}{2\pi \alpha'}\int dr d\tau \h g_{\tau\tau}^{1/2}g_{rr}^{1/2}\sqrt{1+\frac{g_{xx}}{g_{rr}}x'^2}~,
    \label{eq:NGaction}
\end{aligned}
\end{equation}
where primes denote derivatives with respect to $r$.
Since the Lagrangian depends on $x$ only through its derivative, there is a conserved quantity,
\begin{equation}
    \frac{\partial \mathcal{L}}{\partial x'}=\sqrt{\frac{g_{\tau\tau}}{g_{rr}}}\frac{g_{xx}x'}{\sqrt{1+\frac{g_{xx}}{g_{rr}}x'^2}}=\text{const.}\equiv C~.
\end{equation}
This constant can be evaluated at the turning point $r=r_{min}$, where the derivative $x'$ diverges, yielding
\begin{equation}
    C=\sqrt{g_{\tau\tau}g_{xx}}\big\rvert_{r_{min}}~
    \label{eq:C}.
\end{equation}
The resulting equation of motion is
\begin{equation}
    \left(\frac{dx}{dr}\right)^2=C^2 \frac{g_{rr}}{g_{xx}}\frac{1}{g_{\tau\tau}g_{xx}-C^2}~.
\end{equation}
For the interpolating metric \eqref{eq:intermetric}, the equation of motion reduces to
\begin{equation}
    \left(\frac{dx}{dr}\right)^2=\frac{r_{min}^4L^4}{(r^4-r_0^4)(r^4-r_{min}^4)}\left(1-\frac{r_0^4}{r_{min}^4}\cos^2\theta\right)~.
\end{equation}
The separation between the quark and antiquark is then given by
\begin{equation}
    d=2\int_{r_{min}}^\infty |x'(r)|dr=\frac{2L^2}{r_0}\sqrt{a^4-\cos^2\theta}\int_a^\infty\frac{dy}{\sqrt{(y^4-1)(y^4-a^4)}}~,
    \label{eq:dnumerics}
\end{equation}
where we defined $a\equiv r_{min}/r_0$, and rescaled the radial coordinate as $y\equiv r/r_0$. This integral can be expressed in terms of complete elliptic integrals of the first kind \cite{Rey:1998bq},
\begin{equation}
    d=\frac{L^2}{2 r_0\sqrt{a^3 \gamma}}\sqrt{a^4-\cos^2{\theta}}\left(K\left[\sqrt{\frac{\gamma+1}{2\gamma}}\right]-K\left[\sqrt{\frac{\gamma-1}{2\gamma}}\right]\right),
    \label{eq:danalytic}
\end{equation}
with $\gamma\equiv\frac{1}{2}(a+\frac{1}{a})$, and $K(m)$ is the complete elliptic integral of the first kind,
\begin{equation}
    K(k)=\int_0^{\pi/2}\frac{d\theta}{\sqrt{1-k^2\sin^2\theta}}~.
\end{equation}

\begin{figure}
\centering
\begin{minipage}[c]{0.47\textwidth}
    \centering
    \includegraphics[width=\linewidth]{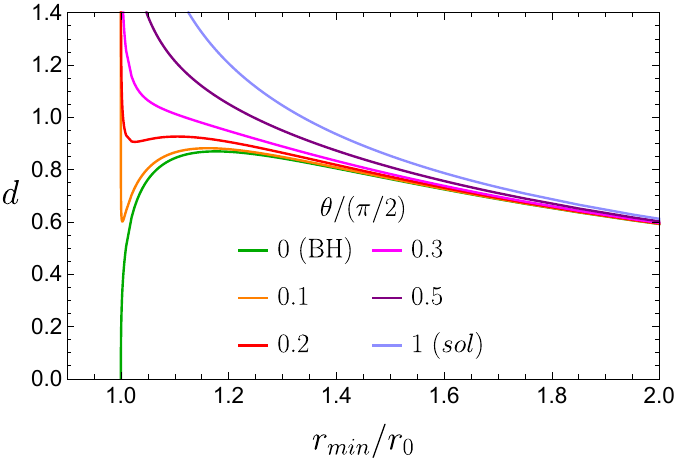}
\end{minipage}
\begin{minipage}[c]{0.52\textwidth}
    \centering
    \includegraphics[width=\linewidth]{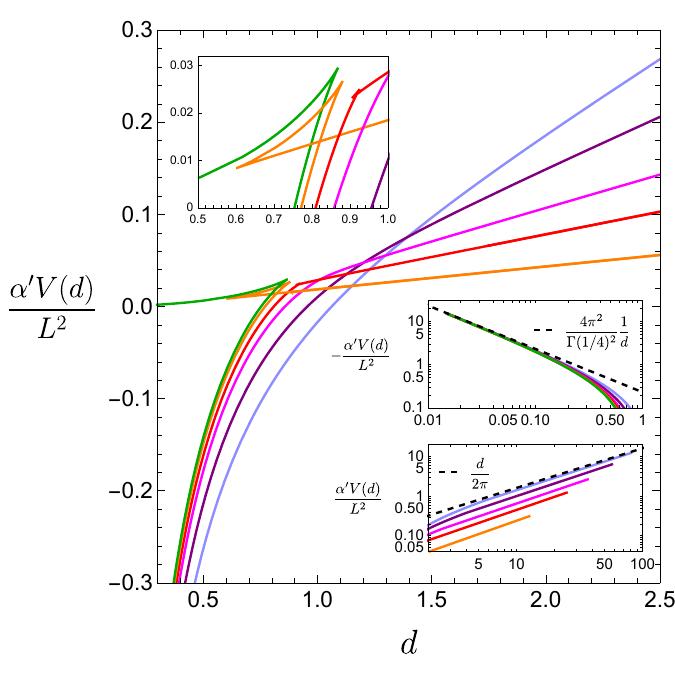}
\end{minipage}
\caption{Numerical solutions for Wilson loops in the geometries with varying $\theta$. Left: quark-antiquark separation $d$ against the minimum radial depth the U-shaped configurations dip to (in units of $r_0$). The limits $\theta\rightarrow 0$ are the black hole and $\theta\rightarrow \pi/2$ the soliton and for each we recover the known results. Right: we show the quark-antiquark potential $V(d)$ for the same angular values as in the left figure. The inset on the top shows a zoom to the near-transition region. The insets on the bottom show in logarithmic scale the small (large) $d$ limits showing the expected Coulomb (linear) behavior.}
\label{fig:distance}
\end{figure}

To extract the quark–antiquark potential we evaluate the on-shell Nambu–Goto action. The potential is obtained from the action as
\begin{equation}
    V(d)=\frac{S_{NG}}{T}\, ,
\end{equation}
where $T=\int d\tau$. As usual, this quantity diverges because the string endpoints extend to the boundary and correspond to infinitely heavy quarks. We therefore regularize the result by subtracting the action of two straight strings stretching from the boundary to the IR. The renormalized potential is thus
\begin{equation}
    V(d)=\frac{1}{\pi\alpha'}\left[
    \int_{r_{min}}^\infty dr\,
    \sqrt{g_{\tau\tau}g_{rr}}
    \left(
    \sqrt{1+\frac{g_{xx}}{g_{rr}}x'^2}
    \right)
    -\int_{r_0}^{\infty} dr\,\sqrt{g_{\tau\tau}g_{rr}}
    \right],
\end{equation}
where the first term corresponds to the U-shaped string configuration and the second term accounts for the subtraction of the two straight strings, making the potential finite.

The resulting quark separation $d$ as a function of $r_{min}/r_0$, together with the interaction potential $V(d)$ are displayed in Figure~\ref{fig:distance} for several values of $\theta$. In all cases, the small-$d$ regime corresponds to to configurations probing only the UV region, where the geometry approaches that of AdS$_5$ and therefore the potential is that of the UV ${\cal N}=4$ SYM, which all our theories share. This is a potential that falls as $-1/d$.

In the left panel, this regime appears on the right-hand side of the curves, while in the right panel it corresponds to the lower portion of the curves at small $d$. To make this behavior more explicit, we include a log–log inset in the right panel focusing on this region. It is then useful to first consider the two limiting cases $\theta=\pi/2$ (soliton) and $\theta=0$ (black hole).

In blue we show the solutions for the soliton geometry ($\theta=\pi/2$). At short distances, the strings probe only the asymptotically AdS region of the geometry, and the potential exhibits the expected Coulombic behavior $V(d)\sim -1/d$. They smoothly map into solutions in which the string reaches the IR wall at $r=r_0$ and develops a segment that lies along this minimal radius. This produces a linear growth of the potential at large $d$, signaling confinement. Note the potential energy passes through zero before entering the linear growth regime. 

At the opposite extreme, the black hole geometry ($\theta=0$) in green, exhibits qualitatively different behavior. At short separations the potential again follows a Coulombic $-1/d$ behavior. However, for sufficiently large separation the preferred configuration is no longer the connected U-shaped string but instead two disconnected strings that fall into the horizon (there is a vertical line at $r_{min}/r_0=1$ not shown on the left plot - so as not to obscure the other lines that merge with it for infinite separation - that represent these solutions). There is no potential for widely separated quarks. 
The black hole solutions shown on the top left in the left hand plot correspond to the potential maxima between the deconfined disconnected strings and the Coulomb solutions. There is a region with three solutions at a given $d$ (including the vertical line not on the left hand plot) which is the swallow tail region around a first order transition. At short separations the Coulomb law applies, whilst at large separations the quarks are deconfined.

For intermediate values of $\theta$, the behavior interpolates smoothly between these limits: as one moves away from the black hole geometry, the solutions that follow the green line (\eg, the orange line) join to a set of solutions that are very close to the disconnected solution of the black hole - they go vertically upwards in the left-hand plot. These are solutions with two almost straight strings tied by a string along $r_0$ of the metric. Since they emerge smoothly from the disconnected solution we know that the string tension grows from zero as $\theta$ grows from $0$. At short separations the potential retains its Coulombic form. Whereas for the black hole there is a first order transition to a deconfined state for growing separation, here there is a first order transition to a small string tension. As $\theta$ increases further, the first order transition becomes less and less pronounced before disappearing to a smooth transition from Coulomb to confined around $0.25 \pi/2$. 

The conclusion is that, except for the black hole geometry at $\theta=0$, the theory always exhibits confinement at sufficiently large quark separation. The confining string tension grows continuously as one moves from the black hole towards the soliton geometry.

The string tension can be measured by simply lying a string along the $r_0$ line representing the mid-section of the string between two infinitely separated quarks. The action is then,
\begin{equation}
    S = \frac{\sqrt{g_{\tau\tau}g_{xx}}|_{r_0}}{2 \pi \alpha'} \int  d\tau dx
\end{equation}
and hence the QCD string tension is controlled by
\begin{equation} \label{eq:stringten}
    T_{QCD} \sim  \sqrt{g_{\tau\tau}g_{xx}} \sim ~r_0^2 \sqrt{1- \cos^2 \theta}.
\end{equation}
The tension vanishes only for the black hole geometry. Thus, for fixed compactification radius $R_z$, varying $\theta$ effectively tunes the temperature and provides a smooth trajectory between the soliton and black hole phases. Although the correspondence between $\theta$ and the temperature is somewhat counterintuitive (placing the black hole at $T=0$ and the soliton at $T\to\infty$), the physical conclusion is clear: confinement disappears continuously and only at the endpoint of the deconfined black hole phase. The string tension changes smoothly and switches off at a continuous transition at $\theta=0$ ($T=0$).

\section{Chiral symmetry breaking}\label{sec:cSB}

Confinement is often expected to induce mass generation through a simple uncertainty principle argument: localization in position leads to higher uncertainty in the momentum. To test the connection between confinement and chiral symmetry breaking in the present setting, we introduce probe D5-branes into the backgrounds described above \cite{Karch:2005ms}. These probes add fundamental matter localized on a $(2+1)$-dimensional defect in the $(3+1)$-dimensional gauge theory.

We consider D5 branes placed at a fixed position in the compact $z$ direction and extended in the rest of the space-time directions $(t, x, y)$, as well as three of the other directions, as summarized in Table~\ref{tab:Embedding structure of the D5-branes and D3-branes}.We are working in a Euclidean space, so our D5 branes are Euclidean also - accounted for by the minus sign in their action.  The D3-D5 open strings correspond to quarks living on the (2+1)-dimensional defect in the 4D gauge theory. Our goal is to determine whether these quarks dynamically acquire a mass, signaling chiral symmetry breaking, as a function of the interpolation parameter $\theta$ in the geometry \eqref{eq:intermetric}.

\begin{table}[h]
    \centering
    \begin{tabular}{|c|c|c|c|c|c|c|c|c|c|c|}
    \hline
       &$0$  &$1$&$2$&$3$&$4$&$5$&$6$&$7$&$8$&$9$  \\
       \hline
       D3 &$-$ &$-$ &$-$ &$(-)$ & $\bullet$ & $\bullet$ & $\bullet$ & $\bullet$ & $\bullet$ &$\bullet$\\
      D5 & $-$ & $-$  & $-$  & $\bullet$ & $-$  & $-$  & $-$  &  $\bullet$ &  $\bullet$ &$\bullet$\\
      \hline
       
    \end{tabular}
    \caption{Embedding structure of the D5-branes and D3-branes.}
    \label{tab:Embedding structure of the D5-branes and D3-branes}
\end{table}

It is convenient to introduce the "isotropic" radial coordinate $u$, defined as \cite{Babington:2003vm}
\begin{equation}
    u^2=\frac{r^2}{2}\left(1+\sqrt{1-\frac{r_0^4}{r^4}}\right)~,\label{eq:isotropiccoord}
\end{equation}
in which the coordinate singularity is now at $u_0=r_0/\sqrt{2}$. The effect of this change of coordinates is to remove the blackening factor from the $g_{rr}$ component of the metric, making the presence of a flat 6-plane perpendicular to the gauge theory coordinates explicitly manifest.

In the new coordinates, the background metric is
\begin{equation}
\begin{aligned}
ds^2=&\frac{u^4+u_0^4}{L^2u^2}\left(1-\frac{4u_0^4u^4}{(u^4+u_0^4)^2}\cos^2\theta\right)d\tau'^2+\frac{u^4+u_0^4}{L^2u^2}(dx^2+dy^2)+\frac{4u_0^4u^2\sin(2\theta)}{L^2(u^4+u_0^4)}d\tau'~dz'\\
    +&\frac{(u^4-u_0^4)^2}{L^2u^2(u^4-u_0^4)^2}\left(1+\frac{4u_0^4u^4}{(u^4-u_0^4)^2}\cos^2\theta\right)dz'^2+\frac{L^2}{u^2}du^2+ L^2d\Omega_5^2~.
\end{aligned}
\end{equation}
A useful parametrization of the 5-sphere is
\begin{equation}
\begin{aligned}
    d\Omega_5^2&=d\varphi^2+\cos^2\varphi d\Omega_2^2+\sin^2\varphi d\tilde{\Omega}_2^2\\
    &=\frac{d\psi^2}{1-\psi^2}+(1-\psi^2)d\Omega_2^2+\psi^2 d\tilde{\Omega}_2^2~,
    \label{eq:metric5spheretheta}
\end{aligned}
\end{equation}
where $\phi\in [0,\pi/2]$, and in the second line we have introduced $\psi\equiv \sin\varphi$ ($\psi\in [0,1]$).

The D5-brane spans the $(\tau',x,y,u,\Omega_2)$, and is located at a fixed position in $\phi$ and the second 2-sphere, $\tilde{\Omega}_2$.  Using an angular embedding $\psi(u)$, the induced metric is
\begin{equation}
\begin{aligned}
    ds^2=&\frac{u^4+u_0^4}{L^2u^2}\left(1-\frac{4u_0^4u^4\cos^2\theta}{(u^4+u_0^4)^2}\right)d\tau'^2+\frac{u^4+u_0^4}{L^2u^2}(dx^2+dy^2)+L^2\left(\frac{1}{u^2}+\frac{\psi'^2}{1-\psi^2}\right)du^2\\
    &+L^2(1-\psi^2)d\Omega_2^2~,
    \label{eq:inducedpsi}
\end{aligned}
\end{equation}

For numerical convenience and as a consistency check, we also introduce \textit{cartesian} coordinates $(\rho,w)$, related to $(u,\phi)$ as
\begin{equation}
    \rho=u\cos\varphi~,\qquad w=u\sin\varphi~,
    \label{eq:relationcartesianangular}
\end{equation}
which satisfy
\begin{equation}
    d\rho^2+dw^2=du^2+u^2d\varphi^2~.
\end{equation}
With an embedding function of the form $w(\rho)$, the induced metric on the D5-brane is 
\begin{equation}
    ds^2=\frac{u^4+u_0^4}{L^2u^2}\left(1-\frac{4u_0^4u^4\cos^2\theta}{(u^4+u_0^4)^2}\right)d\tau'^2+\frac{u^4+u_0^4}{L^2u^2}(dx^2+dy^2)+\frac{L^2}{u^2}\left(1+w'^2\right)d\rho^2+\frac{L^2\rho^2}{u^2}d\Omega_2^2~,
\end{equation}
where $u$ has to be understood as written in terms of the embedding function, $w(\rho)$, as $u^2=\rho^2+w(\rho)^2$.

\begin{figure}
    \centering
    \includegraphics[width=0.5\linewidth]{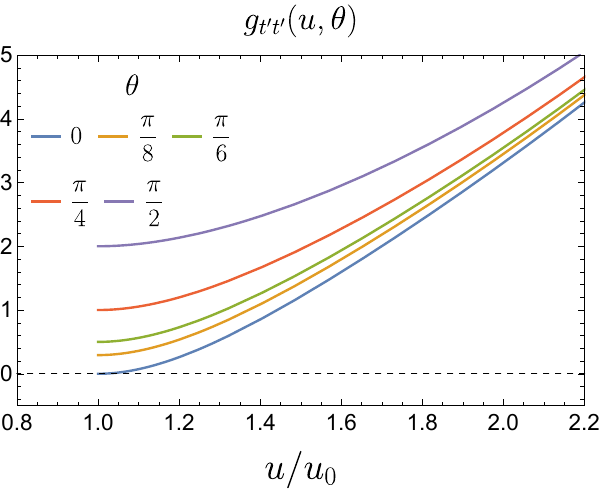}
    \caption{The $g_{\tau'\tau'}$ component of the induced metric. Only in the strict BH limit ($\theta=0$) does the induced metric develop a worldvolume horizon.}
    \label{fig:gttinduced}
\end{figure}

An important point to note is that the induced metric on the D5-brane can develop a worldvolume horizon if $g_{\tau'\tau'}=0$. As seen in Fig.~\ref{fig:gttinduced}, this happens only in the strict black hole limit ($\theta=0$), where the zero coincides with the background black hole horizon. Therefore, at $\theta=0$ both black hole embeddings (where the D5 brane enters the horizon) and Minkowski embeddings exist. However, for any non-zero $\theta$, $g_{\tau'\tau'}$ remains strictly positive, and only Minkowski embeddings exist. It will be interesting to see below how the two types of solutions smoothly transition.

The dynamics of the D5-branes is governed by the DBI action,
\begin{equation}
    S_{D5}=-N_f T_{D5}\int d^6\xi \sqrt{-\det h_{ab}}~,
\end{equation}
where the Wess-Zumino term does not contribute. In angular and cartesian coordinates, the action becomes, respectively,
\begin{equation}
    S_{D5}=-\mathcal{N}\int du\sqrt{\left(1+\frac{u_0^4}{u^4}\right)\left[u^4\left(1+\frac{u_0^4}{u^4}\right)^2-4u_0^4\cos^2\theta\right]}\sqrt{1-\psi^2}\sqrt{1-\psi^2+u^2\psi'^2}
    \label{eq:DBIangular}
\end{equation}
and
\begin{equation}
    S_{D5}=-\mathcal{N}\int d\rho \frac{\rho^2}{(\rho^2+w^2)^3}\left(u_0^4+(\rho^2+w^2)^2\right)^{3/2}\sqrt{1-\frac{4u_0^4\cos^2\theta(\rho^2+w^2)^2}{(u_0^4+(\rho^2+w^2)^2)^2}}\sqrt{1+w'^2}
    \label{eq:DBIw}
\end{equation}
where we have absorbed the prefactors into a constant $\mathcal{N}$, given by
\begin{equation}
    \mathcal{N}=4\pi N_f T_{D5}=\frac{N_fN_c\sqrt{\lambda}}{2\pi^3}~,
\end{equation}
and we have implicitly divided by the (infinite) volume of $\mathbb{R}^{(1,2)}$.
 
Note generically the equations of motion for the embedding should be solved subject to 
\begin{equation}
\partial_\rho w(\rho=0)=0\label{eq:partialrhow}
\end{equation}
these are the so called Minkowski embeddings.
There is only an alternative for the precise black hole case as we discuss next in angular coordinates.

The equation of motion arising from the action \eqref{eq:DBIangular} is of the form
\begin{equation}
    \psi''(u)+\frac{F(u,u_0,\theta,\psi,\psi')}{u^2(u^4+u_0^4)\left(u^8+u_0^8-2u^4u_0^4\cos(2\theta)\right)(1-\psi(u)^2)}=0~,
\end{equation}
where the function $F$ in the numerator is a complicated function which does not play a role in the following discussion. Since $\psi(u)<1$ for black hole embeddings, the equation of motion can only become divergent if $u^8+u_0^8-2u^4u_0^4\cos(2\theta)=0$. This is the same function that makes $g_{\tau'\tau'}$ vanish only for $\theta=0$. In that limit, imposing regularity of the equation of motion at $u=u_0$ gives the IR boundary condition for the integration of black hole embeddings.

For general $\theta$, Minkowski embeddings can be expanded near $u=u_{min}>u_0$ as
\begin{equation}
    \psi(u)=1-\frac{3(u_0^4+u_{min}^4)\left((u_0^8+u_{min}^8-2u_0^4u_{min}^4\cos(2\theta)\right)}{u_{min}(u_{min}^4-u_0^4)\left(3u_0^8+3u_{min}^8+4u_0^4u_{min}^4-2u_0^4u_{min}^4\cos(2\theta)\right)}(u-u_{min})+...
    \label{eq:seriesMink}
\end{equation}
The behavior of $\psi(u)$ near the boundary is (writing until the first order where $\theta$ appears explicitly)
\begin{equation}
    \psi(u)=\frac{\psi_1}{u}+\frac{\psi_2}{u^2}+\frac{\psi_1^2\psi_2}{2u^4}+\frac{\left(6\psi_2^2+u_0^4(4\cos^2\theta-3)\right)\psi_1}{6u^5}+...
    \label{eq:UVasympt}
\end{equation}
The constant $\psi_1$ is identified with the quark mass and $\psi_2$ is the operator sourced by the quark mass that includes the quark condensate \cite{Karch:2005ms}. These constants implicitly depend on $\theta$. 

\subsection{Black Hole Embeddings at \texorpdfstring{$\theta=0$}{theta=0}}\label{cap_Black_Hole_Embeddings}

One limit of our geometry at $\theta=0$ is the black hole case. In this limit, regularity of the equations at the horizon impose the following behavior for $\psi(u)$:
\begin{equation}
    \psi(u)=\psi_0-\frac{\psi_0}{2u_0^2}(u-u_0)^2+...
\end{equation}

\begin{figure}
    \centering
    \includegraphics[width=0.7\linewidth]{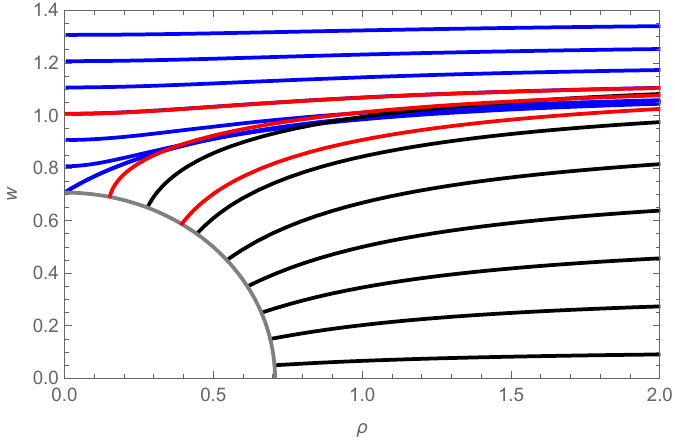}
    \caption{The D5~brane embeddings in the black hole background. All black lines end in the black hole. The blue lines go to $\rho=0$ and never go inside the black hole. There is a first order phase transition where the solutions overlap with a  critical mass  at $m_{Crit}/u_0\simeq1.155$ - the red lines are the embeddings that approach the critical mass.}
    \label{fig:D5_BH}
\end{figure}

We can now find the full set of D5 embeddings by numerically shooting from the horizon or $\rho=0$. We display the well-known solutions for the black hole metric in Figure~\ref{fig:D5_BH} for $u_0=1$. These have the same phase structure as the D7~brane embeddings discussed in \cite{Babington:2003vm,Apreda:2005yz,Mateos:2006nu}.

The only massless embedding is $w=0$, which terminates on the horizon. There is therefore a continuous "meson melting" transition \cite{Hoyos-Badajoz:2006dzi} for a massless quark as soon as temperature is switched on - here bound states acquire an imaginary part to their mass as they become quasi-normal modes of the black hole. 

Embeddings with non-zero quark mass divide into black hole embeddings at low mass and Minkowski embeddings at high mass (where we compare the mass to the temperature). In between there is a first order transition that can be seen from the overlapping solutions in Figure~\ref{fig:D5_BH} that are well known to form a swallow tail first order free energy plot. The critical mass at which the free energies exchange dominance is at around $m/u_0=1.155$. Thus as temperature rises there is a first order meson melting transition for the massive quarks. 

In the theory with compact $z$ direction and temperature, the black hole describes the high temperature phase above the deconfinement transition. The $z$ periodicity, that is not evident in the geometry, plays no role in the D5 physics in this phase. 

\begin{figure}
    \centering    \includegraphics[width=0.7\linewidth]{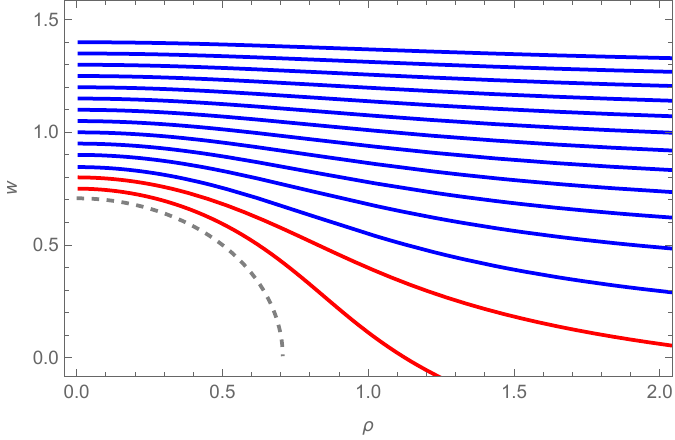}
    \caption{D5~brane embeddings in the soliton background. All red lines go to $w=0$ at a finite $\rho$ and therefore are exited solutions for the D5~branes. The blue lines go to infinity and are real ground state solutions. The starting value of $w$ for which the solutions have the least positive mass is $w(0)\simeq0.8459$. This solution is shown as the lowest blue line.}
    \label{fig:D5_Cig}
\end{figure}

\subsection{The soliton at \texorpdfstring{$\theta=\pi/2$}{theta=pi/2}}

The opposite extreme of our geometry is the soliton (\ref{eq:solitonmetric}). Here, as we have argued, there can only be Minkowski embeddings of the D5 branes. We find them numerically and display the solutions in Figure~\ref{fig:D5_Cig}. There is a single solution for each UV mass value, so there are no transitions. The massless UV solution bends away from $w=0$ in the IR. This leads to a non-zero condensate and a finite $w$ mass-gap in the IR. This is chiral symmetry breaking induced by the conformal symmetry breaking at strong coupling at the scale of the inverse compactification radius. Note in Figure~\ref{fig:D5_Cig} we also show some solutions (in red) that lie closer to the capping off point of the geometry in the IR but extend to negative masses in the UV. These are excited states of the vacuum (one can flip $w \rightarrow -w$) with higher energy than the blue solutions (they still respect the physical IR and UV boundary conditions) - see for example \cite{Filev:2007gb} where they were first discussed.

\begin{figure}
    \centering

 \includegraphics[width=0.32\linewidth]{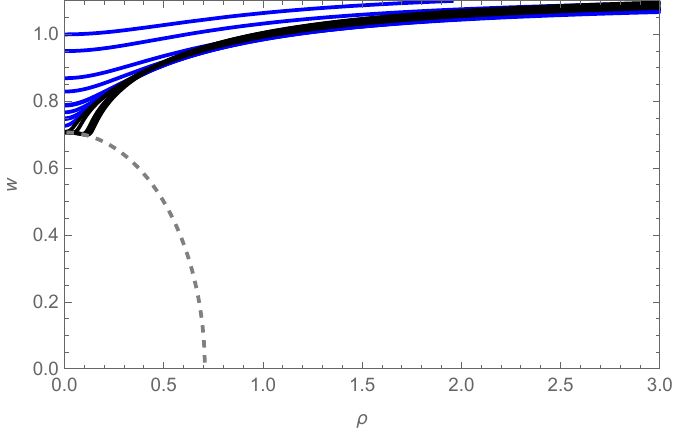}
 \includegraphics[width=0.32\linewidth]{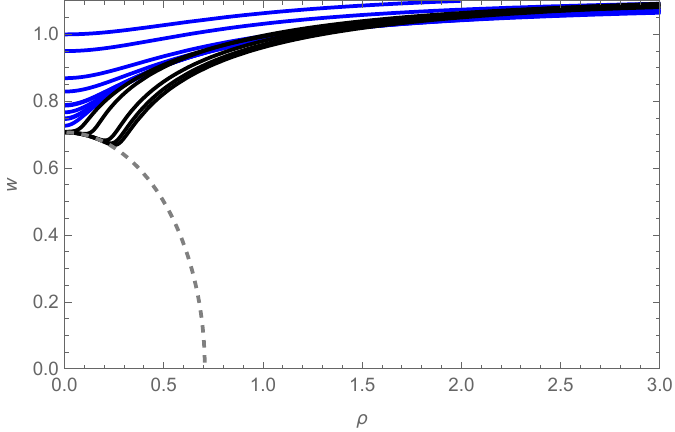}
 \includegraphics[width=0.32\linewidth]{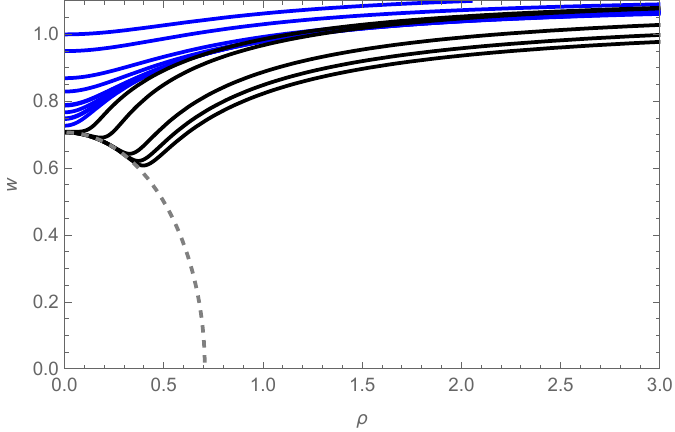}
 \includegraphics[width=0.32\linewidth]{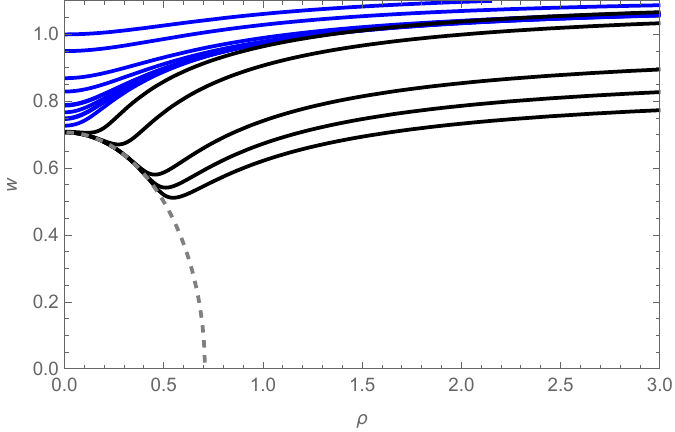}
 \includegraphics[width=0.32\linewidth]{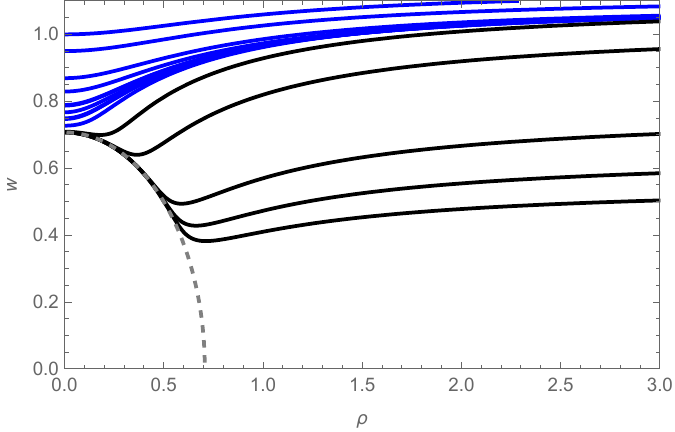}
 \includegraphics[width=0.32\linewidth]{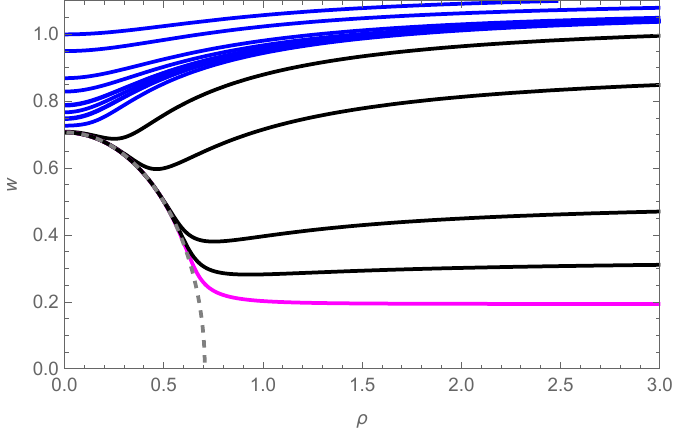}
 \includegraphics[width=0.32\linewidth]{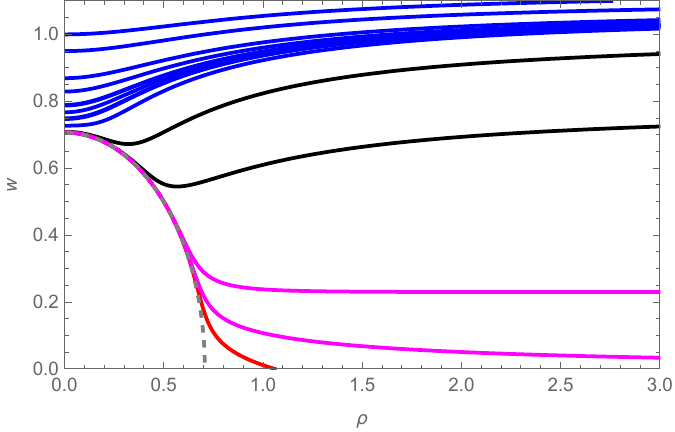}
 \includegraphics[width=0.32\linewidth]{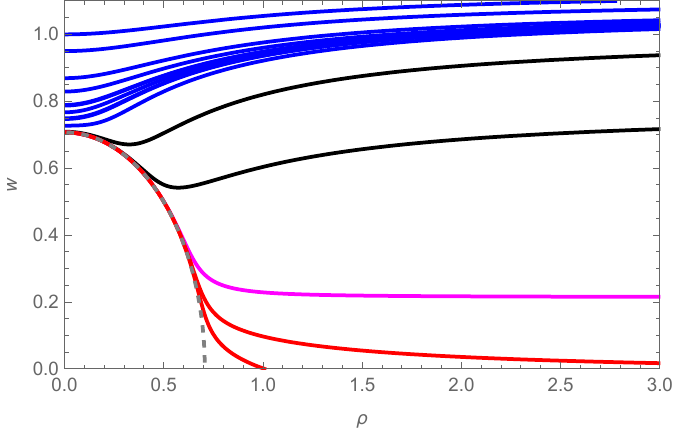}
  \includegraphics[width=0.32\linewidth]{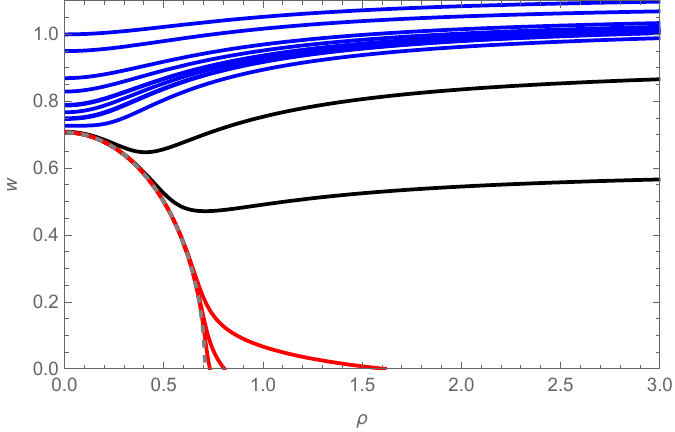}
    \caption{D5~branes in the interpolating backgrounds for $L=r_0=1$ for angles $\theta\in[\frac{\pi}{128},\frac{4\pi}{64}]$ in steps of $\frac{\pi}{128}$ (from top left to bottom right). The seventh graph is for $\theta=\frac{3.4725\pi}{64}$. The starting conditions at $\rho=0$ of the embeddings are kept fixed in each case. The red solutions are excited states of the D5 brane since they go to $w=0$ at a finite $\rho$. The magenta lines start near the cap of the geometry at $u_0=1/\sqrt{2}$, then curve down and do not curve up for greater $\rho$. In the $\rho\to\infty$ limit they either have positive or 0 prefactor in the $\frac{1}{\rho}$ term. The black lines have a minimum in $w$ at some $\rho$. The blue lines never curve down and always increase.  }
    \label{fig:D5_CS_breaking_Mix}
\end{figure}

\subsection{Interpolating geometry at intermediate \texorpdfstring{$\theta$}{theta}  }

The geometries in (\ref{eq:intermetric}) interpolate between the black hole and the soliton metrics. The first interesting question is how the transition from the black hole D5 embedding solutions in Figure~\ref{fig:D5_BH} transition to the Minkowski embeddings of Figure~\ref{fig:D5_Cig}. As we have seen above the interpolating metric can only have Minkowski solutions at all non-zero $\theta$.  To understand the transition we first set $r_0=1$ so we can study the embeddings at different $\theta$. Note here one could simply consider the base geometry with a compact radius $r_0$ and rotate the D5 brane embedding coordinates to achieve the same results. This set up is a preliminary to considering the variation in behavior with $T$ where one must also vary the diagonal length in Fig 1. We will return to that full problem at the end of this sub-section.

We show some sample embeddings, found numerically, in Figure~\ref{fig:D5_CS_breaking_Mix}. Here at fixed $L=1, r_0=1$we vary $\theta$ and show a set of embeddings which in each figure begin with the same IR value of the embeddings $L(\rho=0)$. We see how embeddings that lie a little above the cap take over the role of the black hole embeddings at $\theta=0$.  
The transition is most easily understood by considering the bottom three frames which are for slightly larger $\theta$ values but where it is easier to find the Minowski embeddings that asymptote towards zero mass in the UV. All the low mass embeddings are of Minkowski type ending just slightly above $r_0$ at $\rho=0$. They then wrap the "cap" of the geometry (which is very low cost in action - turning to zero cost at $\theta=0$ when it becomes an horizon) before exiting at some point to infinite $\rho$. As one approaches $\theta=0$, the embeddings lie on the horizon before emerging to match Figure~\ref{fig:D5_BH}. 

In the other images in Figure~\ref{fig:D5_CS_breaking_Mix}, we show the evolution of the first order transition at finite mass (associated with the region where there are multiple overlapping solutions describing a first order swallow tail in free energy). At $\theta=0$ this transition is usually associated with the appearance of quasi-normal modes on the black hole embedding, but at finite $\theta$ the remnant transition is always between Minkowski embeddings and there are not quasi-normal modes - simply a jump in the quark condensate and the meson masses.  By the final bottom right image the overlapping functions have ceased and the first order transition has terminated at a fixed point. The critical point where the transition terminates is at $\theta= 3.4725 \pi/64$.

An alternative approach to understand the chiral symmetry breaking instability in our family of geometries is to explicitly show the instability of the $w=0$ solution in the backgrounds.  Generically, the action for a dimension one field, $\hat{w}$, in AdS$_4$  is
\begin{equation} 
S \sim \int d^3x ~d \rho ~\left( \frac{1}{2} \rho^2 (\partial_\rho \hat{w})^2 + \frac{1}{2}  M^2 \hat{w}^2 \right)  \label{eq:BFform}
 \end{equation}
Note expanding the action (\ref{eq:DBIw}) around $w=0$ at large $\rho$ gives precisely this action with $M^2=0$. 
The Breitenlohner-Freedman bound would be violated were $M^2 < -1/4$ (here the equation of motion has solution $w=\rho^{-1/2}$ and both operator and source have dimension $d/2$). 

 We can place the full action (\ref{eq:DBIw}) near $w=0$ in this form. First expanding we have (dropping the vacuum energy)
\begin{equation}
    S = \int d^3x ~d\rho  \left(\frac{1}{2} A(\rho) ( \partial_\rho w)^2 + \frac{1}{2} B(\rho) w^2 \right)
\end{equation}
where $A,B$ are functions of $\rho$.
To place this in the form  of (\ref{eq:BFform}) we must redefine the radial coordinate so that the kinetic term is canonical \cite{Alvares:2012kr}. We have
\begin{equation}
    A(\rho)  \partial_\rho = \tilde{\rho}^2 \partial_{\tilde{\rho}}, \hspace{1cm} \tilde{\rho} = \frac{1}{ \int_\rho^\infty \frac{d \rho}{A(\rho)}}~,
\end{equation}
and the action becomes
\begin{equation}
    S = \int d^3x ~d\tilde{\rho} \left(\frac{1}{2}\tilde{\rho}^2 (\partial_{\tilde{\rho}} w)^2 + \frac{1}{2 \tilde{\rho}^2 }A(\rho)B(\rho) w^2 \right)~.
        \end{equation} 

    To be able to see if the BF bound is violated, it is sufficient to plot this mass squared, $M^2(\rho) = A(\rho)B(\rho) /\rho^2$, against $\rho$ (note not against $\tilde{\rho}$ which is numerically more complicated and has $\theta$ dependent ranges).  

\begin{figure}
    \centering    \includegraphics[width=0.55\linewidth]{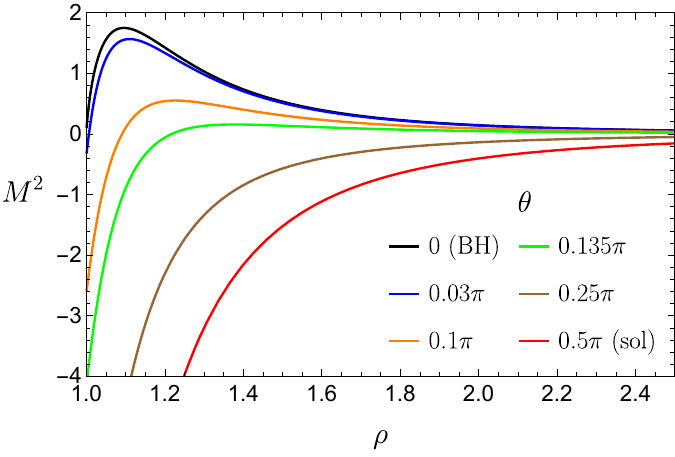}
    \caption{Effective mass squared, $M^2(\rho) = A(\rho)B(\rho) /\rho^2$, of the D5 embedding field in the chiral symmetry preserving configuration $w=0$ as a function of the radial coordinate $\rho$ for several values of $\theta$ at fixed $u_0=1$. The BF bound is violated, triggering chiral symmetry breaking, if $M^2$ passes through $-1/4$, which occurs before the geometry singularity for all cases except for $\theta=0$ (and a very narrow neighborhood around it).}
    \label{fig:M2_D5}
\end{figure}

We plot the mass squared as a function of the radial coordinate $\rho$ for several values of $\theta$ in Figure~\ref{fig:M2_D5}. For the cases with $\theta >0$, the BF bound is violated (the mass squared fall below $-1/4$). This signals an instability of the $\hat{w}=0$ configuration, indicating that the true ground state will lie off axis and have chiral symmetry breaking present. By contrast, in the strict BH limit $\theta=0$, the BF bound is not violated outside the horizon and the $\hat{w}=0$ solution is stable. The evolution is smooth, consistent with our observation of a continuous transition described above.

We note, however, that there is a small region very close to $\theta=0$ (see for example the $\theta=0.03 \pi$ case)  where this plot suggests the BF bound is only violated at scales below $u_0$, yet it is violated above the scale of the singularity in (\ref{eq:BFform}) - it seems likely the solutions react to the lower scale BF bound violation to remain smooth. These values of $\theta$ lie too close to the BH limit for our numerical analysis to totally confirm the behavior (the solutions lie on the horizon essentially until they hit the $L=0$ solution).

\begin{figure}
    \centering
    \includegraphics[width=0.7\linewidth]{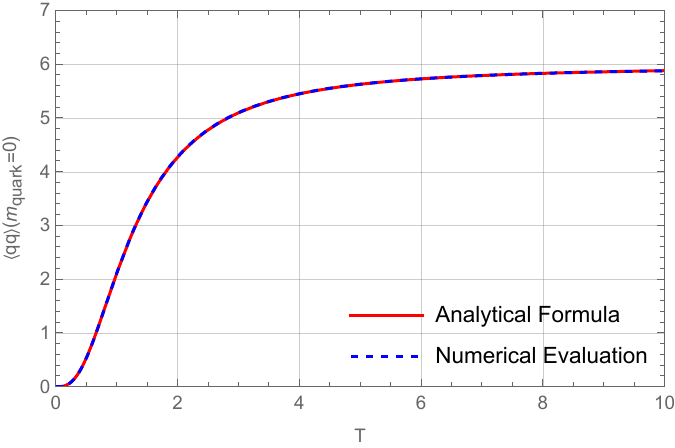}
    \caption{The quark condensate $\langle q\overline{q}\rangle|_{m=0}$ as one tracks with temperature along the unstable maxima of the swallow tail shown in Figure~\ref{fig:freeenergy}. The continuous switch off of the condensate at $T=0$ is apparent. We also plot the curve $0.604 r_0^2 \sqrt{1-(\cos{\theta})^2}=0.604\pi^2(\frac{T^2}{1+T^2})^{3/2}$ to show the condensate is very tightly linked to the string tension in (\ref{eq:stringten}) , where the prefactor is chosen as $\langle q\overline{q}\rangle|_{m=0}$ for $\theta =\pi/2$}.
    \label{fig:condensate}
\end{figure}

So far in this subsection we have varied $\theta$ at fixed $r_0$ to see chiral symmetry breaking switch off smoothly as one approaches $\theta=0$. To link this analysis back to our original problem where we fixed $R_z$ and varied the radius of the time coordinate to vary temperature we must appropriately rescale $r_0$. 
From Eq. (2.11), rearranging prefactors we have
    \begin{equation}
        r_0=\frac{L^2}{2R_z}\frac{1}{\sqrt{1+\frac{1}{4\pi^2R_z^2T^2}}}\,.
    \end{equation}
    For our numerical work we set $L=1$ and $R_z=\frac{1}{2\pi}$ (as we did for the free energy plots), then  we are left with $\theta\to \arctan{T}$ and 
    \begin{equation}
        r_0=\frac{\pi}{\sqrt{1+\frac{1}{T^2}}}\,.
    \end{equation}
    
We then track along the unstable branch in Figure~\ref{fig:freeenergy}. Note that the unstable branch tracks from a black hole solution at $T=0$ to a soliton solution at infinite $T$.

As an example, we  look at the variation of the quark condensate expectation value for massless quarks ($\langle q\overline{q}\rangle|_{m=0}$) with $T$. We evaluate $\langle q\overline{q}\rangle|_{m=0}$ from the sub-leading UV form of the solution for the D5 embedding as in (\ref{eq:UVasympt}). We show the results in Fig.~\ref{fig:condensate}. One can see that $\langle q\overline{q}\rangle|_{m=0}$ smoothly switches off as  $T\to 0$ and also that $\langle q\overline{q}\rangle|_{m=0}$ asymptotes to a fixed value for $T\rightarrow \infty$, matching the value in the soliton case. Therefore, the interpolating metric has a continuous transition in $\langle q\overline{q}\rangle|_{m=0}$ with $T$. Moreover, since the quark condensate has dimension 2 in the $d=2+1$ defect theory, if it is dynamically driven by confinement one expects it to exhibit the same temperature scaling as the string tension in \eqref{eq:stringten}. We overlay this scaling behavior in Fig.~\ref{fig:condensate}, finding excellent agreement with the numerical results. This strongly suggests this model has a single scale for both chiral symmetry breaking and confinement.

\section{Discussion}

In this paper we have revisited the thermal deconfinement transition of ${\cal N}=4$ SYM compactified on a spatial circle, with the goal of better understanding the relation between confinement and chiral symmetry breaking in a holographic setting. Our central construction is a one-parameter family of Euclidean geometries that completes the familiar swallow-tail structure of the free energy associated with the first-order transition. The new saddle corresponds to the unstable local maximum of the effective potential, interpolating continuously between the two stable extrema: the AdS soliton and the AdS black hole. Geometrically, this configuration is characterized by making manifest the identification of diagonally separated corners of the $\tau$–$z$ boundary torus. In the free energy diagram shown in Figure~\ref{fig:freeenergy}, this saddle appears as the dotted branch connecting the two minima.

Although this geometry is thermodynamically unstable, it is precisely this linking of the  minima that makes it useful for our purposes. It provides a continuous path through the space of bulk solutions connecting a confining phase and a deconfined phase of the dual gauge theory. Along this path, the roles of temperature and geometry are somewhat counterintuitive: the confining soliton corresponds to the low-temperature endpoint, while the deconfined black hole arises at high temperature, and the unstable branch interpolates between them in the manner dictated by the swallow-tail structure. By probing this branch with strings and D5 branes, we have been able to track confinement and chiral symmetry breaking continuously across the transition.

Confinement was diagnosed through the string tension extracted from Wilson loops, computed via the standard U-shaped fundamental string configurations. At short distances, the ${\cal N}=4$ theory has a Coulomb interaction between quarks. In the soliton geometry, at large separations the string settles near the capped-off IR region, producing a linear confining potential. As we smoothly decrease temperature on the unstable branch, the string tension decreases smoothly. At zero temperature the string tension vanishes, where the string lies along the cap that has become the black hole horizon and the solutions smoothly map on to the usual finite temperature results. In particular there is, on this unstable branch,  a continuous transition with quark separation to disconnected, screened quark states. Confinement therefore smoothly switches off as $T \rightarrow 0$. This behavior is summarized in our Figure~\ref{fig:distance}.

To study chiral symmetry breaking we included D5 probe branes in the geometries (introducing quarks in the non compact directions of the geometry). In the soliton geometry, the IR cap repels the D5 brane, leading to dynamical mass generation and chiral symmetry breaking even for vanishing UV quark mass. Along the unstable branch, a crucial feature is that only Minkowski embeddings exist: the D5 branes do not end on the cap except in the strict black hole limit, where the cap corresponds to the black hole horizon. As the temperature decreases, solutions for low quark mass begin to wrap near the cap with the action cost for the piece near the cap decreasing smoothly. In the $T\to 0$ limit along the unstable branch, these portions of the embedding lie precisely on the surface that becomes the black hole horizon, and the solutions smoothly connect to the familiar black hole embeddings in which the D5 branes fall through the horizon. As a result, the chiral condensate decreases continuously and vanishes only at the deconfined endpoint, as illustrated in Figure~\ref{fig:condensate}. Chiral symmetry breaking here seems to be directly linked to confinement (see the fit in Figure \ref{fig:condensate}) and one should presumably infer that the confinement causes the chiral symmetry breaking.

All of this analysis is an interesting story in its own right. Beyond these results, our broader motivation was to gain insight into the microscopic origin of confinement in holography. We had wondered at what point along the smooth transition confinement and chiral symmetry breaking  would switch off and whether we could correlate this behavior with expectation values of gauge-invariant operators. Our analysis suggests a natural interpretation in terms of the boundary stress-energy tensor. The expectation value of $T^{\tau\tau}$, associated with the subleading behavior of $g^{\tau\tau}$ \cite{Haehl:2012tw}, characterizes the presence of a thermal bath in the deconfined phase. In contrast, in the soliton geometry $T^{\tau\tau}$ vanishes, but there is a pressure $T^{zz}$ and confinement. The $T^{zz}$ vev is therefore the natural  operator to represent a gauge invariant description of a vacuum density of monopoles. One might have anticipated that this quantity could vanish at some intermediate point along the unstable branch, for example at $\theta=\pi/4$, and confinement and chiral symmetry breaking be lost tying those phenomena to that vev. Instead, we find that $T^{zz}$ only vanishes precisely at horizon formation. This indicates a more intricate interplay between monopole density and the thermal bath density, but nonetheless reinforces the idea that confinement is tied to the persistence of this pressure-like contribution.

There are several natural directions in which this work could be extended. One obvious generalization is to introduce finite baryon density via a worldvolume gauge field on the D5 branes. In models without confinement, such a density is known to suppress chiral symmetry breaking \cite{Evans:2010hi}, but since a probe cannot modify the capping-off of the background geometry, it is unclear whether a similar effect would arise here. We note that at finite temperature and density, brane models may exhibit tachyonic modes leading to instabilities \cite{Kaminski:2009ce}. A further promising avenue is to include an R-charge chemical potential, leading to a charged (Reissner–Nordström) black hole background, and to study how this additional control parameter influences confinement and chiral symmetry breaking. We hope that the framework developed here will be useful for exploring these and related questions.

\acknowledgments

We are grateful to Gunnar Bali for discussions.
J.E., N.E.~and~F.V.~acknowledge funding through the DFG Research Training Group `Particle physics at colliders in the LHC precision era' (GRK2994).
N.E.’s work was supported by the STFC consolidated grant ST/X000583/1. The work of M.B. has been funded by Xunta de Galicia through the Programa de axudas \'a etapa predoutoral and the grant 2023-PG083 with reference code ED431F 2023/19.

\bibliographystyle{JHEP}
\bibliography{biblio.bib}

\end{document}